\documentclass[
 reprint,
 amsmath,amssymb,
 aps,
 superscriptaddress
]{revtex4-2}

\usepackage{graphicx}
\usepackage{hyperref}
\usepackage[linesnumbered, ruled, vlined, titlenotnumbered, noend]{algorithm2e}
\usepackage{amsmath, amsfonts, amssymb, amsthm, color}
\usepackage{multirow} 
\usepackage{tikz-cd}
\usepackage[capitalise]{cleveref}
\usepackage{tikz}
\usepackage{subcaption}
\usetikzlibrary{shapes.geometric, arrows.meta, positioning, fit}
\crefname{algocf}{alg.}{algs.}
\Crefname{algocf}{Algorithm}{Algorithms}

\newcommand{\ket}[1]{|#1\rangle}

\newcommand{\Z}{{\mathbb{Z}}}

\newcommand{\polyA}{{\cal A}}
\newcommand{\polyB}{{\cal B}}

\newcommand{\datamodule}{M^{d}}
\newcommand{\ancillamodule}{M^{a}}
\newcommand{\loss}{\text{Loss}}

\tikzset{
    module/.style={
        rectangle,
        rounded corners=1.5mm,
        minimum width=1.8cm,
        minimum height=0.5cm,
        draw=black,
        line width=0.5pt
    },
    ion/.style={
        circle,
        fill=black,
        inner sep=1pt,
        minimum size=2.5mm,
        draw=black,
        line width=0.5pt
    },
    interconnect/.style={
        draw=blue!70!cyan,
        line width=2.5pt,
        -{Stealth[length=3mm, width=3mm]},
    },
    cross/.style={
        red,
        line width=2pt
    },
    merged/.style={ 
        draw=black,
        line width=1.5pt,
        dashed,
        rounded corners=1.5mm,
        inner sep=0.7mm 
    }
}

\usepackage{lipsum}

\newcommand\ploss{\ensuremath{p_{\text{loss}} } }

\begin{document}

\author{Nolan J. Coble}
\affiliation{
    IonQ Inc.
}
\affiliation{Department of CS and Joint Center for Quantum Information and Computer Science (QuICS), University of Maryland, College Park, MD 20742}

\author{Min Ye}
\affiliation{
    IonQ Inc.
}

\author{Nicolas Delfosse}
\affiliation{
    IonQ Inc.
}

\title{Correction of chain losses in trapped ion quantum computers}

\date{\today}

\begin{abstract}
Neutral atom quantum computers and to a lesser extent trapped ions may suffer from atom loss.
In this work, we investigate the impact of atom loss in long chains of trapped ions.
Even though this is a relatively rare event, ion loss in long chains must be addressed because it destabilizes the entire chain resulting in the loss of all the qubits of the chain.
We propose a solution to the chain loss problem based on 
(1) a quantum error correction code distributed over multiple long chains, 
(2) beacon qubits within each long chain to detect the loss of a chain,
and 
(3) a decoder adapted to correct a combination of circuit faults and erasures after beacon qubits convert chain losses into erasures.
We verify the chain loss correction capability of our scheme through circuit level simulations with a distributed $[[72,12,6]]$ BB code with beacon qubits.
\end{abstract}

\maketitle

\section{Introduction}

Trapped ions are one of the most promising platforms for quantum computing thanks to their record fidelity~\cite{hughes2025trapped} and the ability to move ions, which provides a great flexibility in terms of architecture design as illustrated by the QCCD architecture~\cite{kielpinski2002architecture}.

To reach large-scale applications, a fault-tolerant architecture capable of correcting faults during a computation is necessary~\cite{shor1996fault}. 
Faults induced by imperfect qubit operations are corrected using a quantum error correction code such as surface codes in the MUSIQC architecture~\cite{monroe2014large} or quantum LDPC codes adapted to trapped ions in~\cite{ye2025quantum, tham2025distributed}.

Another important source of noise is the loss of an ion, {\em i.e.}, a qubit, which refers to the physical disappearance of the ion. This typically results from heating or collisions with residual background gas molecules in the vacuum chamber~\cite{bruzewicz2019trapped, de2021materials, strohm2024ion}.
Even though the loss rate of trapped ions is typically lower than technologies like neutral atoms~\cite{evered2023high} (an ion can be trapped for days or even months in some cases~\cite{gabrielse1990thousandfold, bruzewicz2019trapped}), ion losses cannot be ignored because they spread via qubit interactions. If left uncontrolled, the loss of an ion might rapidly propagate to a large fraction of the qubits during a computation.
One can deal with ion losses using erasure correction codes~\cite{vala2005quantum} combined with loss detection units (originally introduced as a leakage detection in ~\cite{preskill1998fault}, see also~\cite{perrin2025quantum}), or by replacing measurements by state-selective measurements that distinguish between the three outcomes 0, 1 and loss, together with loss post-selection~\cite{reichardt2024fault} or a delayed erasure decoder~\cite{baranes2025leveraging}.
Fast reloading is also required to replace the lost ions~\cite{bruzewicz2016scalable}; see also neutral atom reloading~\cite{li2025fast, chiu2025continuous}.

These approaches immediately apply to a trapped ion architecture based on ions confined in their own potential well, and merged during two-qubit gates as in the QCCD architecture~\cite{kielpinski2002architecture}.
However, the problem of ion loss was not considered in the case of long ion chains where a large number of ions are confined in the same potential well, despite the popularity of long chains in quantum computing experiments. Chains with five ions are demonstrated in~\cite{monz2016realization, debnath2016demonstration}, chains with 10 to 20 ions in~\cite{monz201114, nam2020ground, egan2021fault, egan2021scaling, postler2022demonstration}, a long chain forming a 30-qubit quantum computer is benchmarked in~\cite{chen2024benchmarking}, and analog quantum simulations using chains with up to 53 ions are performed in~\cite{zhang2017observation, kranzl2022controlling}.

The problem of ion loss in long chains is not only unexplored, it is also significantly more challenging than in the neutral atom or short chain case because the loss of an ion is likely to induce the loss or complete destabilization of the entire chain, resulting in the simultaneous loss of many qubits at once. Indeed, Table 4.1 of \cite{vittorini2013stability}, presents experimental data showing that for chains up to six ions, a single ion loss usually results in the entire chain being lost, or only one ion remaining. Even if some ions remain, the loss instantly breaks the delicate force balance between the trap potential and Coulomb repulsion. The remaining ions are forced to rapidly re-equilibrate, resulting in a global shift of their positions and injecting significant motional heat into the system's collective modes. This positional shift has immediate consequences for measurement: since the detection lasers are precisely aimed at the original ion locations, the displaced ions will fall outside the laser focus. Consequently, the measurement system will register no emitted photons, which suggests the ion is in the $\ket{0}$ (dark) state when, in fact, the qubit information is simply lost due to misaligned optics.

In this work, we propose a solution to the ion loss problem in long chains of trapped ions based on three ingredients that we further describe in the next paragraphs:
(1) quantum error correction codes distributed over multiple long chains
(2) beacon qubits within each long chain to detect the chain loss and
(3) a decoder correcting a combination of circuit faults and chain losses.

Distributing the code's data qubits over multiple long chains is necessary because it is impossible to recover for the loss of all the data qubits.
This excludes quantum error correction codes supported over a single long chain as in the QEC experiments reported in~\cite{egan2021fault, egan2021scaling, postler2022demonstration} or the protocols designed in~\cite{ye2025quantum}.
The distributed quantum error correction protocols from~\cite{tham2025distributed} seem more robust against chain losses because each chain contains only a fraction of the data qubits. For example, the [[144,12,12]] bivariate bicycle (BB) code of~\cite{bravyi2024high} is distributed over 12 long chains containing 12 data qubits each.
However, when distributing a code, the data qubits contained in each chain must be selected carefully to make sure that none of the logical operators is fully supported inside a chain. Otherwise, the loss of this chain, would permanently delete some of the logical information.
In this work, we refine the distribution of the BB codes to avoid this issue and to make distributed BB codes robust against chain loss.

Within each ion chain, we use one or a few qubits, that we call {\em beacon qubits}, to certify that the chain is still present.
The basic idea is to use the fact that the outcome 1 during the measurement of a qubit corresponds to the detection of photons emitted by the corresponding ion. If a chain is lost, its ions cannot emit photons anymore and all measurements return the outcome 0.
Therefore, to verify that a chain was not lost, we use beacon qubits, kept in state $\ket 1$, and measured at regular time intervals. As long as the chain is present, the beacon measurement should return the outcome 1. If a beacon returns the outcome 0, it is likely that the chain is lost.
This strategy might fail to immediately detect a chain loss in the presence of circuit faults because a fault might flip an outcome 0, hiding a chain loss.
To make sure that chain losses are immediately detected, we use multiple beacon qubits per chain. We expect that a few beacons are sufficient to suppress the probability of an undetected chain loss below the logical error rate so that this phenomenon does not significantly affect the code performance. Alternatively, one can use a single beacon qubit per chain and repeat its measurement.

Once a chain loss is detected, it can be replaced by a new chain from a reservoir that is continuously reloaded~\cite{bruzewicz2016scalable, li2025fast, chiu2025continuous}.
The qubits of the replacement chain are initialized in the maximally mixed state, which effectively converts the chain loss into an erasure and we adapt the decoder to deal with a combination of circuit faults and erasure using standard techniques~\cite{delfosse2021almost, wu2022erasure, kubica2023erasure, kang2023quantum, gu2025fault}.
The main difference with~\cite{baranes2025leveraging} is that chain losses are immediately detected after each round of circuit operations thanks to our beacon qubits, removing the need for delayed erasure decoding.

Though our protocol differs from previous ideas in that it is distributive, it is compatible with some earlier strategies. 
For instance, one could replace our beacon qubits with the loss detection units proposed in \cite{preskill1998fault, perrin2025quantum}.
We prefer the beacon qubit because it does not consume any two-qubit gate.
Our protocol is also compatible with the use of state-selective measurements for ancilla measurements~\cite{reichardt2024fault, baranes2025leveraging}. We prefer the introduction of a beacon qubit that is constantly measured to guarantee an earlier detection of chain losses.
However, one could envision combining both approaches.

We perform numerical simulations showing the efficacy of our distributed quantum error correction protocol against circuit faults and chain losses.
Our simulations demonstrate that the scheme mitigates these loss events, achieving logical error rates close to the no-chain-loss baseline.

In \cref{sect:review model}, we review the distributed quantum computing model of ~\cite{tham2025distributed} and we extend it to include chain losses.
The distributed quantum error correction scheme of \cite{tham2025distributed} is adapted to tolerate chain losses in \cref{sect:beacon qubits} by introducing beacon qubits and by adjusting the mapping of the data qubits onto chains and the syndrome extraction circuit.
Our numerical results are presented in \cref{sect:numerical results}.
We conclude and discuss several future directions in~\cref{sect:conclusion}.

\section{The $2\times L$ model with chain loss}
\label{sect:review model}

We begin with a summary of the $2\times L$ module array model proposed in \cite{tham2025distributed}.

\begin{itemize}
    \item \textbf{Structure and Rows:} The architecture consists of a $2 \times L$ array of cells labeled $(b, i) \in \mathbb{Z}_2 \times \mathbb{Z}_L$, where each cell contains a module (a register of $n$ qubits). Cells $(0, i), i\in\Z_L$ form the \textbf{fixed row}, and Cells $(1, i), i\in\Z_L$ form the \textbf{moving row}.
    
    \item \textbf{Operations:} Two-qubit unitary gates are supported either inside a module or in a pair of \textbf{aligned modules} (cells $(0, i)$ and $(1, i)$). Operations acting on different modules can be performed \textbf{simultaneously}.
    
    \item \textbf{Cyclic Shift:} The key mechanism for connectivity is the \textbf{cyclic shift} (or $s$-shift), which moves all modules in the moving row, transporting the module in cell $(1, i)$ to cell $(1, (i+s) \mod L)$.
    
    \item \textbf{Shift Depth:} Any cyclic shift is assumed to have \textbf{depth one}, meaning the shift duration is independent of the shift size $s$.
\end{itemize}

The \textbf{Sparse Cyclic Layout} (Algorithm~2 in \cite{tham2025distributed}) is a quantum circuit designed to efficiently perform syndrome extraction for bivariate bicycle (BB) codes \cite{bravyi2024high} within this architecture. It produces a \textbf{short-depth syndrome extraction circuit} by strategically interleaving cyclic shifts and highly parallel two-qubit gates across different modules.
In the next section, we will present a slight variation of Algorithm 2 from \cite{tham2025distributed}. The primary difference is that each module in our layout contains fewer data qubits. This modification ensures that the number of lost data qubits in a single chain loss remains below the code distance, as exceeding the code distance would render the chain loss unrecoverable.

Given that the loss of a single ion destabilizes the entire ion chain, our model treats this as a complete chain-loss event. We integrate this phenomenon into the $2 \times L$ module array model by assuming that each chain is lost independently at each time step with probability $\ploss$. This probabilistic model is detailed further in Appendix~\ref{sect:noise model}.

\begin{figure}
\begin{subfigure}{\linewidth}
\begin{tikzpicture}
    \node[module, label={[yshift=0.4mm]above:{\footnotesize D0}}] (module-top-1) {};
    \foreach \i in {1,...,4} {
        \node[ion] at ([xshift={-0.6cm + (\i-1)*0.4cm}]module-top-1.center) {};
    }

    \node[module, label={[yshift=0.4mm]above:{\footnotesize D1}}, right=0.5cm of module-top-1] (module-top-2) {};
    \foreach \i in {1,...,4} {
        \node[ion] at ([xshift={-0.6cm + (\i-1)*0.4cm}]module-top-2.center) {};
    }

    \node[module, label={[yshift=0.4mm]above:{\footnotesize D2}}, right=0.5cm of module-top-2] (module-top-3) {};
    \foreach \i in {1,...,4} {
        \node[ion] at ([xshift={-0.6cm + (\i-1)*0.4cm}]module-top-3.center) {};
    }

    \node[module, label=above:{\footnotesize A0}, below=0.5cm of module-top-1] (module-bottom-1) {};
    \foreach \i in {1,...,4} {
        \node[ion] at ([xshift={-0.6cm + (\i-1)*0.4cm}]module-bottom-1.center) {};
    }

    \node[module, label=above:{\footnotesize A1}, right=0.5cm of module-bottom-1] (module-bottom-2) {};
    \foreach \i in {1,...,4} {
        \node[ion] at ([xshift={-0.6cm + (\i-1)*0.4cm}]module-bottom-2.center) {};
    }

    \node[module, label=above:{\footnotesize A2}, right=0.5cm of module-bottom-2] (module-bottom-3) {};
    \foreach \i in {1,...,4} {
        \node[ion] at ([xshift={-0.6cm + (\i-1)*0.4cm}]module-bottom-3.center) {};
    }

    \draw[cross] (module-bottom-1.north west) -- (module-bottom-1.south east);
    \draw[cross] (module-bottom-1.north east) -- (module-bottom-1.south west);

\end{tikzpicture}
\caption{Ancilla module A0 is lost.}
\end{subfigure}

\begin{subfigure}{\linewidth}
\begin{tikzpicture}
    \node[module, label={[yshift=0.4mm]above:{\footnotesize D0}}] (module-top-1) {};
    \foreach \i in {1,...,4} {
        \node[ion] at ([xshift={-0.6cm + (\i-1)*0.4cm}]module-top-1.center) {};
    }

    \node[module, label={[yshift=0.4mm]above:{\footnotesize D1}}, right=0.5cm of module-top-1] (module-top-2) {};
    \foreach \i in {1,...,4} {
        \node[ion] at ([xshift={-0.6cm + (\i-1)*0.4cm}]module-top-2.center) {};
    }

    \node[module, label={[yshift=0.4mm]above:{\footnotesize D2}}, right=0.5cm of module-top-2] (module-top-3) {};
    \foreach \i in {1,...,4} {
        \node[ion] at ([xshift={-0.6cm + (\i-1)*0.4cm}]module-top-3.center) {};
    }

    \node[module, label=above:{\footnotesize A0}, below=0.5cm of module-top-1] (module-bottom-1) {};
    \foreach \i in {1,...,4} {
        \node[ion] at ([xshift={-0.6cm + (\i-1)*0.4cm}]module-bottom-1.center) {};
    }

    \node[module, label=above:{\footnotesize A1}, right=0.5cm of module-bottom-1] (module-bottom-2) {};
    \foreach \i in {1,...,4} {
        \node[ion] at ([xshift={-0.6cm + (\i-1)*0.4cm}]module-bottom-2.center) {};
    }

    \node[module, label=above:{\footnotesize A2}, right=0.5cm of module-bottom-2] (module-bottom-3) {};
    \foreach \i in {1,...,4} {
        \node[ion] at ([xshift={-0.6cm + (\i-1)*0.4cm}]module-bottom-3.center) {};
    }


    \draw[interconnect] ([yshift=0cm]module-bottom-1.east) -- ([yshift=0cm]module-bottom-2.west);
    
    \draw[interconnect] ([yshift=0cm]module-bottom-2.east) -- ([yshift=0cm]module-bottom-3.west);

    \draw[interconnect] ([yshift=0cm]module-bottom-3.east)
        -- ++(0.5,0)
        arc (0:-90:0.5) 
        -- ++(-6.6,0) 
        arc (270:180:0.5) 
        -- ([yshift=0cm]module-bottom-1.west);

    \draw[cross] (module-top-1.north west) -- (module-top-1.south east);
    \draw[cross] (module-top-1.north east) -- (module-top-1.south west);
    \draw[cross] (module-bottom-1.north west) -- (module-bottom-1.south east);
    \draw[cross] (module-bottom-1.north east) -- (module-bottom-1.south west);

    \node[merged, fit=(module-top-1) (module-bottom-1)] {};
    \node[merged, fit=(module-top-2) (module-bottom-2)] {};
    \node[merged, fit=(module-top-3) (module-bottom-3)] {};
\end{tikzpicture}
\caption{Upon merging aligned modules, both A0 and D0 are lost.}
\end{subfigure}

\begin{subfigure}{\linewidth}
\begin{tikzpicture}
    \node[module, label={[yshift=0.4mm]above:{\footnotesize D0}}] (module-top-1) {};
    \foreach \i in {1,...,4} {
        \node[ion] at ([xshift={-0.6cm + (\i-1)*0.4cm}]module-top-1.center) {};
    }

    \node[module, label={[yshift=0.4mm]above:{\footnotesize D1}}, right=0.5cm of module-top-1] (module-top-2) {};
    \foreach \i in {1,...,4} {
        \node[ion] at ([xshift={-0.6cm + (\i-1)*0.4cm}]module-top-2.center) {};
    }

    \node[module, label={[yshift=0.4mm]above:{\footnotesize D2}}, right=0.5cm of module-top-2] (module-top-3) {};
    \foreach \i in {1,...,4} {
        \node[ion] at ([xshift={-0.6cm + (\i-1)*0.4cm}]module-top-3.center) {};
    }

    \node[module, label=above:{\footnotesize A2}, below=0.5cm of module-top-1] (module-bottom-1) {};
    \foreach \i in {1,...,4} {
        \node[ion] at ([xshift={-0.6cm + (\i-1)*0.4cm}]module-bottom-1.center) {};
    }

    \node[module, label=above:{\footnotesize A0}, right=0.5cm of module-bottom-1] (module-bottom-2) {};
    \foreach \i in {1,...,4} {
        \node[ion] at ([xshift={-0.6cm + (\i-1)*0.4cm}]module-bottom-2.center) {};
    }

    \node[module, label=above:{\footnotesize A1}, right=0.5cm of module-bottom-2] (module-bottom-3) {};
    \foreach \i in {1,...,4} {
        \node[ion] at ([xshift={-0.6cm + (\i-1)*0.4cm}]module-bottom-3.center) {};
    }


    \draw[interconnect] ([yshift=0cm]module-bottom-1.east) -- ([yshift=0cm]module-bottom-2.west);
    
    \draw[interconnect] ([yshift=0cm]module-bottom-2.east) -- ([yshift=0cm]module-bottom-3.west);

    \draw[interconnect] ([yshift=0cm]module-bottom-3.east)
        -- ++(0.5,0)
        arc (0:-90:0.5) 
        -- ++(-6.6,0) 
        arc (270:180:0.5) 
        -- ([yshift=0cm]module-bottom-1.west);

    \draw[cross] (module-top-1.north west) -- (module-top-1.south east);
    \draw[cross] (module-top-1.north east) -- (module-top-1.south west);
    \draw[cross] (module-bottom-1.north west) -- (module-bottom-1.south east);
    \draw[cross] (module-bottom-1.north east) -- (module-bottom-1.south west);
    \draw[cross] (module-top-2.north west) -- (module-top-2.south east);
    \draw[cross] (module-top-2.north east) -- (module-top-2.south west);
    \draw[cross] (module-bottom-2.north west) -- (module-bottom-2.south east);
    \draw[cross] (module-bottom-2.north east) -- (module-bottom-2.south west);

    \node[merged, fit=(module-top-1) (module-bottom-1)] {};
    \node[merged, fit=(module-top-2) (module-bottom-2)] {};
    \node[merged, fit=(module-top-3) (module-bottom-3)] {};
\end{tikzpicture}
\caption{After a cyclic shift, A2 and D1 are also lost.}
\end{subfigure}

\begin{subfigure}{\linewidth}
\begin{tikzpicture}
    \node[module, label={[yshift=0.4mm]above:{\footnotesize D0}}] (module-top-1) {};
    \foreach \i in {1,...,4} {
        \node[ion] at ([xshift={-0.6cm + (\i-1)*0.4cm}]module-top-1.center) {};
    }

    \node[module, label={[yshift=0.4mm]above:{\footnotesize D1}}, right=0.5cm of module-top-1] (module-top-2) {};
    \foreach \i in {1,...,4} {
        \node[ion] at ([xshift={-0.6cm + (\i-1)*0.4cm}]module-top-2.center) {};
    }

    \node[module, label={[yshift=0.4mm]above:{\footnotesize D2}}, right=0.5cm of module-top-2] (module-top-3) {};
    \foreach \i in {1,...,4} {
        \node[ion] at ([xshift={-0.6cm + (\i-1)*0.4cm}]module-top-3.center) {};
    }

    \node[module, label=above:{\footnotesize A1}, below=0.5cm of module-top-1] (module-bottom-1) {};
    \foreach \i in {1,...,4} {
        \node[ion] at ([xshift={-0.6cm + (\i-1)*0.4cm}]module-bottom-1.center) {};
    }

    \node[module, label=above:{\footnotesize A2}, right=0.5cm of module-bottom-1] (module-bottom-2) {};
    \foreach \i in {1,...,4} {
        \node[ion] at ([xshift={-0.6cm + (\i-1)*0.4cm}]module-bottom-2.center) {};
    }

    \node[module, label=above:{\footnotesize A0}, right=0.5cm of module-bottom-2] (module-bottom-3) {};
    \foreach \i in {1,...,4} {
        \node[ion] at ([xshift={-0.6cm + (\i-1)*0.4cm}]module-bottom-3.center) {};
    }


    \draw[interconnect] ([yshift=0cm]module-bottom-1.east) -- ([yshift=0cm]module-bottom-2.west);
    
    \draw[interconnect] ([yshift=0cm]module-bottom-2.east) -- ([yshift=0cm]module-bottom-3.west);

    \draw[interconnect] ([yshift=0cm]module-bottom-3.east)
        -- ++(0.5,0)
        arc (0:-90:0.5) 
        -- ++(-6.6,0) 
        arc (270:180:0.5) 
        -- ([yshift=0cm]module-bottom-1.west);

    \draw[cross] (module-top-1.north west) -- (module-top-1.south east);
    \draw[cross] (module-top-1.north east) -- (module-top-1.south west);
    \draw[cross] (module-bottom-1.north west) -- (module-bottom-1.south east);
    \draw[cross] (module-bottom-1.north east) -- (module-bottom-1.south west);
    \draw[cross] (module-top-2.north west) -- (module-top-2.south east);
    \draw[cross] (module-top-2.north east) -- (module-top-2.south west);
    \draw[cross] (module-bottom-2.north west) -- (module-bottom-2.south east);
    \draw[cross] (module-bottom-2.north east) -- (module-bottom-2.south west);
    \draw[cross] (module-top-3.north west) -- (module-top-3.south east);
    \draw[cross] (module-top-3.north east) -- (module-top-3.south west);
    \draw[cross] (module-bottom-3.north west) -- (module-bottom-3.south east);
    \draw[cross] (module-bottom-3.north east) -- (module-bottom-3.south west);

    \node[merged, fit=(module-top-1) (module-bottom-1)] {};
    \node[merged, fit=(module-top-2) (module-bottom-2)] {};
    \node[merged, fit=(module-top-3) (module-bottom-3)] {};
\end{tikzpicture}
\caption{After one more cyclic shift, all modules are lost.}
\end{subfigure}

\caption{Catastrophic chain loss: The loss of one module leads to the subsequent loss of all modules after several cyclic shift and merge operations. The top row (D0, D1, D2) contains fixed modules of data qubits, and the bottom row (A0, A1, A2) contains moving modules of ancilla qubits. A dashed rectangle indicates merged modules.}
\label{fig:catastrophic}
\end{figure}
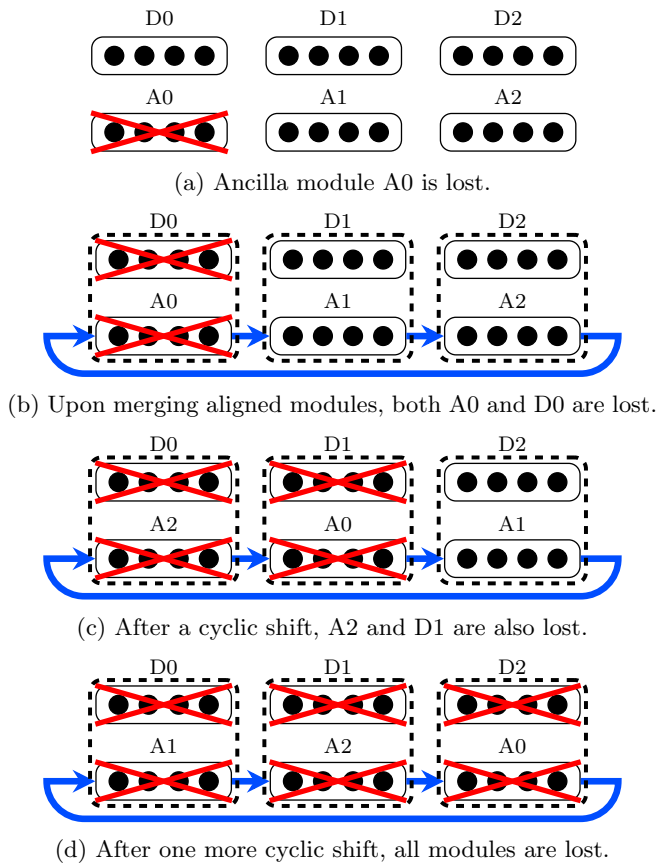

Chain loss is particularly detrimental in a distributed setting, as it can escalate to a catastrophic failure of the entire system. This is because the loss of a single module propagates to other functioning modules during the cyclic shift and merge operations. As this process repeats, the failure spreads rapidly, ultimately resulting in the loss of the entire array and all encoded data, as illustrated in \cref{fig:catastrophic}.

\section{Chain loss correction using Beacon qubits}
\label{sect:beacon qubits}

To detect chain-loss events, we introduce a scheme based on ``beacon qubits". A dedicated beacon qubit is incorporated into each ion chain module, and its sole purpose is to be periodically measured to verify the physical presence and optical alignment of the chain. The protocol operates by initializing the beacon qubit in the $\ket{1}$ (``bright") state, which is defined to scatter photons during measurement. If the measurement detector registers the expected photons, the chain is confirmed to be present. If, however, no photons are detected, the system flags this outcome as a chain-loss event. 

A single beacon qubit is susceptible to two failure modes from circuit faults: {\bf (i) False Positives}: A functional chain is incorrectly flagged as lost. This occurs if the beacon qubit, prepared in the $\ket{1}$ state, is measured as $\ket{0}$. This can be caused by a Pauli $X$ error flipping the qubit to $\ket{0}$ or a measurement error where the detector fails to register the emitted photons. {\bf (ii) False Negatives}: A real chain loss is missed. This occurs if the ``no photons" ($\ket{0}$) signal from a lost chain is incorrectly measured as $\ket{1}$ (a measurement error).

To address both failure modes, we can employ multiple beacon qubits per ion chain instead of a single one. The decision of whether the chain is lost is then based on a majority vote of the $b$ beacon measurement outcomes. For example, if $b=3$, a loss is declared only if two or more beacons return $\ket{0}$. This redundancy provides resilience against both failure modes. A single, random fault (e.g., a measurement error flipping one $\ket{1}$ to $\ket{0}$) is no longer sufficient to cause a false positive, as it will be outvoted by the other two $\ket{1}$ outcomes. Similarly, a single fault hiding a real loss (flipping one $\ket{0}$ to $\ket{1}$) will be outvoted by the other beacons that correctly report $\ket{0}$. This method allows for a significant, tunable reduction in both error rates at the cost of increased qubit overhead.



Upon detection of a chain loss, the system immediately replenishes the lost module with fresh qubits. These new qubits are initialized in the maximally-mixed state $I/2$, which signifies a total loss of information as they possess no correlation with the original encoded state. This physical replacement, combined with the knowledge of the loss event's location from the beacon qubit measurements, is equivalent to an erasure error~\cite{grassl1997codes, bennett1997capacities} which is typically easier to correct than a Pauli error~\cite{delfosse2020linear, chang2024surface, connolly2024fast, gokduman2024erasure}.

While this scheme reliably converts chain loss into erasures, successful correction still requires a careful co-design of the layout and syndrome extraction circuit. For instance, if all data qubits supporting a logical operator reside on a single ion chain, the loss of that chain would be uncorrectable.
We ensure correctability against a single chain loss by imposing the constraint that the number of data qubits per chain must be less than the code distance. This constraint necessitates modifying the syndrome extraction circuit presented in Algorithm 2 of \cite{tham2025distributed}.

We begin by reviewing BB code construction and the sparse cyclic layout from Algorithm 2 of \cite{tham2025distributed}, and then present our modification. BB codes are defined by two integers $\ell$ and $m$, and six integer pairs $(i_1,j_1), \dots,(i_6, j_6) \in \Z_\ell \times \Z_m$. The code has $n=2\ell m$ data qubits and $n/2=\ell m$ stabilizers of both X and Z types. Data qubits are labeled by $(u,v,w) \in \Z_2 \times \Z_\ell \times \Z_m$. The X and Z ancilla qubits are labeled $(X,v,w)$ and $(Z,v,w)$, respectively, for $(v,w) \in \Z_\ell \times \Z_m$. 
An X ancilla qubit $(X,v,w)$ measures the weight-6 X stabilizer supported on $(0, v \oplus i_1, w \oplus j_1), (0, v \oplus i_2, w \oplus j_2), (0, v \oplus i_3, w \oplus j_3), (1,v \oplus i_4, w \oplus j_4), (1,v \oplus i_5, w \oplus j_5), (1,v \oplus i_6, w \oplus j_6)$.
A Z ancilla qubit $(Z,v,w)$ measures the weight-6 Z stabilizer supported on $(0, v \ominus i_4, w \ominus j_4), (0, v \ominus i_5, w \ominus j_5), (0, v \ominus i_6, w \ominus j_6), (1,v \ominus i_1, w \ominus j_1), (1,v \ominus i_2, w \ominus j_2), (1,v \ominus i_3, w \ominus j_3)$. Here, $\oplus$ and $\ominus$ denote addition and subtraction modulo $\ell$ or $m$, respectively.
Note that our description is equivalent to the standard description of BB codes using parity check matrices; see Section~IV of \cite{tham2025distributed} for a proof.

In \cite{tham2025distributed}, the modules are indexed by $w \in \Z_m$. Each data module consists of the data qubits in the set $\datamodule_{w} := \Z_2 \times \Z_\ell \times \{w\}$. Each ancilla module consists of the ancilla qubits in the set $\ancillamodule_{w} := \{X,Z\} \times \Z_\ell \times \{w\}$.
Define two sets $\polyA=\{(i_1,j_1), (i_2,j_2), (i_3,j_3)\}$ and $\polyB=\{(i_4,j_4), (i_5,j_5), (i_6,j_6)\}$.
We use them to introduce shorthand notations for the required sets of CX gates:
\begin{itemize}
\item $\text{CX}[(i,j)\in\polyA]$ refers to the set of CX gates controlled on qubit $(X,v,w)$ targeting qubit $(0, v \oplus i, w \oplus j)$ for all $v\in\Z_\ell$ and $w\in\Z_m$.
\item $\text{CX}[(i,j)\in\polyB]$ refers to the set of CX gates controlled on qubit $(X, v,w)$ targeting qubit $(1,v \oplus i, w \oplus j)$ for all $v\in\Z_\ell$ and $w\in\Z_m$.
\end{itemize}
\cref{algorithm:sparse_cyclic_layout_recap} details the circuit from \cite{tham2025distributed} for measuring the X stabilizers of BB codes.

\begin{algorithm}
\DontPrintSemicolon
\SetAlgoLined
Prepare all the $X$ ancilla qubits in the state $\ket +$.\;
\For{each distinct $j \in \{j_1,j_2,\dots,j_6\}$}{
    Apply the cyclic shift aligning $\ancillamodule_{0}$ and $\datamodule_{j}$.\;
    \For{$i \in \Z_{\ell}$ such that $(i,j)\in\polyA$}{
        Execute all the CX gates in $\text{CX}[(i,j)\in\polyA]$.
    }
    \For{$i \in \Z_{\ell}$ such that $(i,j)\in\polyB$}{
        Execute all the CX gates in $\text{CX}[(i,j)\in\polyB]$.
    }
}
Measure all the ancilla qubits in the $X$ basis.\;
\caption{Recap of Algorithm~2 from \cite{tham2025distributed}}
\label{algorithm:sparse_cyclic_layout_recap}
\end{algorithm}

All simulations in this paper use the $[[72,12,6]]$ BB code from \cite{bravyi2024high}, which is defined by parameters $\ell=m=6$, $\polyA=\{(3,0), (0,1), (0,2)\}$, and $\polyB=\{(0,3), (1,0), (2,0)\}$. For this specific code, the set $\{j_1, \dots, j_6\}$ contains only 4 distinct elements. Consequently, only 4 cyclic shifts are required to measure the X stabilizers.

However, for the $[[72,12,6]]$ BB code, this layout assigns $12$ data qubits per module, significantly exceeding the code distance of $6$ and rendering the loss of any single module uncorrectable.
To address this issue, we modify the layout to assign only 3 qubits to each module. We assume $\ell$ is even, which holds for the $[[72,12,6]]$ code with $\ell=6$. The data modules in our modified layout are indexed by $(r, s, w)\in \Z_2 \times \Z_2 \times \Z_m$ and defined as 
\begin{align*}
& \datamodule_{(r,0,w)} := \{r\} \times \{0,\dots,\ell/2-1\} \times \{w\}, \\
& \datamodule_{(r,1,w)} := \{r\} \times \{\ell/2,\dots,\ell-1\} \times \{w\}.
\end{align*}
The ancilla modules in our modified layout are indexed by $(r, s, w)\in \{X,Z\} \times \Z_2 \times \Z_m$ and defined as 
\begin{align*}
& \ancillamodule_{(r,0,w)} := \{r\} \times \{0,\dots,\ell/2-1\} \times \{w\}, \\
& \ancillamodule_{(r,1,w)} := \{r\} \times \{\ell/2,\dots,\ell-1\} \times \{w\}.
\end{align*}

With this module assignment, one round of X stabilizer measurements for the $[[72,12,6]]$ BB code can be completed using only 6 shifts, as opposed to 4 shifts in the case of 12 qubits per module. The 6 shifts are detailed below. 

The first shift aligns $\ancillamodule_{(X,s,w)}$ with $\datamodule_{(0,1-s,w)}$ for all $s\in\Z_2$ and $w\in\Z_6$. This alignment enables the execution of all CX gates in $\text{CX}[(3,0)\in\polyA]$.

The second shift aligns $\ancillamodule_{(X,s,w)}$ with $\datamodule_{(0,s,w\oplus 1)}$ for all $s\in\Z_2$ and $w\in\Z_6$. This alignment enables the execution of all CX gates in $\text{CX}[(0,1)\in\polyA]$.

The third shift aligns $\ancillamodule_{(X,s,w)}$ with $\datamodule_{(0,s,w\oplus 2)}$ for all $s\in\Z_2$ and $w\in\Z_6$. This alignment enables the execution of all CX gates in $\text{CX}[(0,2)\in\polyA]$.

The 4th shift aligns $\ancillamodule_{(X,s,w)}$ with $\datamodule_{(1,s,w\oplus 3)}$ for all $s\in\Z_2$ and $w\in\Z_6$. This alignment enables the execution of all CX gates in $\text{CX}[(0,3)\in\polyB]$.

The 5th shift aligns $\ancillamodule_{(X,s,w)}$ with $\datamodule_{(1,s,w)}$ for all $s\in\Z_2$ and $w\in\Z_6$.
The 6th shift aligns $\ancillamodule_{(X,s,w)}$ with $\datamodule_{(1,1-s,w)}$ for all $s\in\Z_2$ and $w\in\Z_6$.
These two shifts together enable the execution of all CX gates in $\text{CX}[(1,0)\in\polyB]$ and $\text{CX}[(2,0)\in\polyB]$.

\section{Numerical results}
\label{sect:numerical results}

We perform Monte Carlo simulations using Stim~\cite{gidney2021stim} and employ the BP-OSD decoder~\cite{panteleev2021degenerate, roffe_decoding_2020, Roffe_LDPC_Python_tools_2022} for decoding. Specifically, we use the BP-OSD implementation in \cite{Roffe_LDPC_Python_tools_2022} with 10,000 min-sum BP iterations followed by order-5 combination-sweep OSD. 
We assume the beacon qubits provide perfect detection of chain loss. As previously analyzed, the false positives and false negatives of this mechanism can be exponentially suppressed by increasing the number of beacon qubits per chain. Under our model, this addition does not affect the logical error rate, as all qubit measurements are performed in parallel. The only trade-off is the increased qubit overhead.

\begin{figure}
    \centering
    \includegraphics[width=0.9\linewidth]{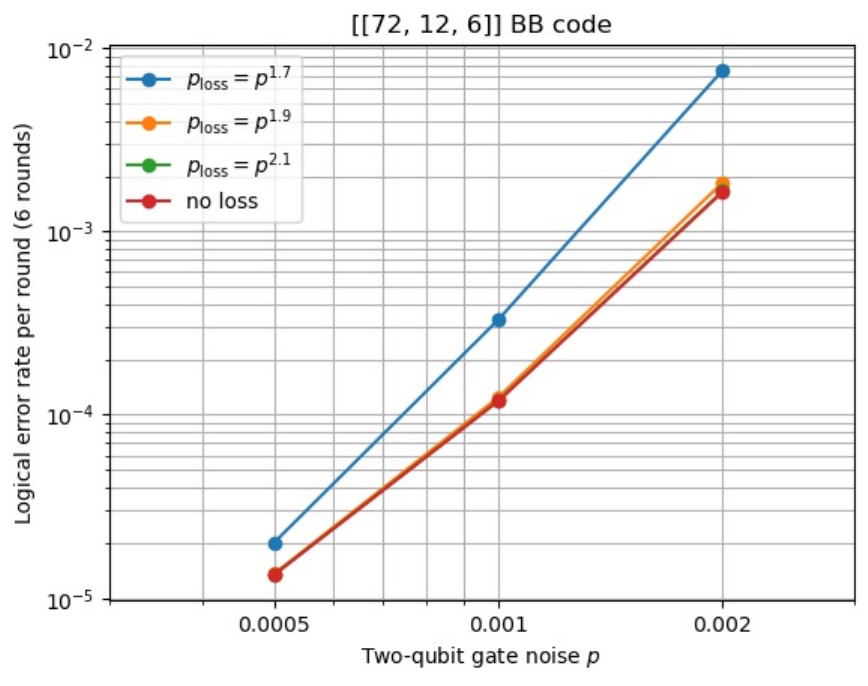}
    \caption{Logical error rate as a function of $p$, with $\ploss = p^\alpha$ for $\alpha \in \{1.7, 1.9, 2.1\}$.}
    \label{fig:with_no_loss_case}
\end{figure}

The noise model used in our simulation is detailed in Appendix~\ref{sect:noise model}. Since Stim does not natively support chain or qubit loss events, we developed a sampling-based method to simulate them, as explained in Appendix~\ref{sect:simulation method}. All simulations in this paper use the $[[72,12,6]]$ BB code from \cite{bravyi2024high}.
In \cref{fig:with_no_loss_case} and \cref{fig: logical error rate physical error}, the ``logical error rate" on the y-axis refers specifically to the X logical error rate.

In \cref{fig:with_no_loss_case}, we plot the logical error rate as a function of the 2-qubit gate error rate $p$, setting the chain loss probability to $\ploss = p^\alpha$ for $\alpha \in \{1.7, 1.9, 2.1\}$. This simulation assumes an ideal scenario where beacon qubit measurements are instantaneous, i.e., they take zero time and induce no idling noise on other qubits. Meanwhile the measurements of other qubits still take time according to the noise model.
Under this assumption, our distributed layout together with the beacon qubit scheme performs extremely well. For low chain loss probabilities $\ploss=p^{1.9}$ and $\ploss=p^{2.1}$, the logical error rate is essentially identical to the no-loss baseline. Even for the relatively high loss rate of $\ploss=p^{1.7}$, the logical error rate at $p=10^{-3}$ remains within a 3x factor of the no-loss case.

In \cref{sect:pessimistic_beacon}, we present simulation results under a more pessimistic beacon qubit measurement model. In this scenario, beacon measurements are no longer instantaneous and are instead assumed to have the same time duration as all other qubit measurements.

In Appendix~\ref{sect:single chain loss}, we investigate the effect of a single chain loss, showing that its impact on the logical error rate is highly dependent on the time step at which it occurs.

\section{Conclusions and future directions}
\label{sect:conclusion}

We proposed a strategy to correct chain loss in trapped ion architecture using a quantum error correction code distributed over multiple long chains and beacon qubits added to each long chain to detect the chain loss.

Our protocol is based on the distributed quantum error correction scheme proposed in~\cite{tham2025distributed}. It would be interesting to adapt other distributed quantum error correction schemes~\cite{sinclair2023fault, de2024thresholds, sutcliffe2025distributed} to the correction of chain losses.

The problem considered in this paper is analogous to the classical use of erasure correction codes in distributed systems, which must deal with nodes that become temporarily or permanently inaccessible. Reed Solomon codes~\cite{reed1960polynomial} have been used in this context for decades~\cite{plank1997tutorial} and are part of Google's file system Colossus~\cite{Ma2012Colossus} and Facebook's file system~\cite{muralidhar2014f4, rashmi2014hitchhiker}, whereas Microsoft's file system uses local reconstruction codes~\cite{huang2012erasure}. 
Another popular choice is network codes~\cite{dimakis2011survey}, recently generalized to the quantum setting~\cite{delfosse2024correction}.
One could consider importing ideas from classical distributed computing to the quantum setting to deal with chain losses.

Our numerical results show the viability of this approach and demonstrate that this protocol benefits from fast beacon qubit measurements. 
It would be interesting to explore the design of such fast measurements, exploiting for instance the fact that beacon qubits never interact with other qubits (through two-qubit gates), which makes them less affected by noise, and could reduce the required cooling time which is a major contribution to the operation time of trapped ion quantum computers~\cite{moses2023race}.

To limit the time overhead of our chain loss correction protocol, we may consider measuring the beacon qubit less often and focusing on the most sensitive locations in the circuit, that is the location where a chain loss is the most likely to induce a logical error. \cref{sect:single chain loss} makes a first step in that direction by  investigating the impact of a chain loss at all possible location of the syndrome circuit.

We focused on the correction of ion loss or chain loss. A related noise source is the leakage error which maps the state of an ion outside of the two levels used to encode a qubit.
Leakage errors are less harmful than ion loss for long chain because they do not trigger a global error affecting the whole chain.
Therefore, we expect that standard leakage detection and correction techniques can be applied~\cite{preskill1998fault, aliferis2005fault, suchara2015leakage, auger2017blueprint, stricker2020experimental, cong2022hardware, miao2023overcoming, chow2024circuit, yu2024processing, yu2025locating}.

For simplicity, we assume that a lost chain can be immediately replaced by another chain from a reservoir regularly reloaded. We leave the design of the reservoir and the chain routing for future work.

\section{Acknowledgment}
The authors thank Jeremy Sage, Aharon Brodutch, Edwin Tham, Felix Tripier, JP Marceaux, Woochang Chung and John Gamble and the whole IonQ team for useful and insightful discussions.

\bibliography{references}

\appendix

\section{Noise model in the simulation}
\label{sect:noise model}

We used the chain model of~\cite{ye2025quantum} as our noise model to simulate qubit operations inside modules.
The model assumes that two-qubit gates are sequential inside a module, but gates acting on distinct modules can be performed simultaneously.
\begin{itemize}
    \item \textbf{Two-qubit gates}: error rate $p$.
    \item \textbf{Single-qubit operations} (including unitary gate, reset, and measurement): error rate $p/10$.
    \item \textbf{Idle qubits}: error rate $p/100$.
    \item \textbf{Measurement duration ($\tau_m$)}: We assume unmeasured qubits undergo $\tau_m = 30$ rounds of idle noise during a measurement operation.
    \item \textbf{Cyclic Shift duration ($\tau_s$)}: A cyclic shift is followed by depolarizing noise on all qubits with a rate of $\tau_s p/100$, where $\tau_s=20$. This means all qubits suffer from $\tau_s$ rounds of idle noise during a shift.
\end{itemize}

Our simulation model assumes all operations (unitary gates, resets, measurements, and cyclic shifts) have depth one in the syndrome extraction circuit, executing in a single time step. In practice, the chain loss probability for a given time step may depend on the specific operations executed, as different operations usually have different durations.
However, we neglect such dependence and simply model chain loss as an independent event occurring at any time step, to any module, with a uniform probability $\ploss=p^\alpha$. Here, $p$ is the two-qubit gate error rate and $\alpha$ is a model parameter. To reflect that chain loss is a rare event relative to gate errors, we set $\alpha > 1$. In our simulations, we specifically investigate $\alpha \in \{1.5, 1.7, 1.9, 2.1\}$, which ensures the chain loss probability $\ploss$ is significantly smaller than $p$.

In the syndrome extraction circuit, data and ancilla modules alternate between merging with each other and standing alone. We assume that if the loss occurs while two modules are merged, the entire merged module (all 3 data and 3 ancilla qubits) is erased to the maximally-mixed state. On the other hand, if the loss occurs during an unmerged step, only the 3 qubits in the lost module are erased. We further assume that the loss probability is independent of module size, meaning a merged data/ancilla module is just as likely to be lost as an unmerged one.

\section{Method for simulating chain loss in Stim}
\label{sect:simulation method}

One difficulty with Stim is that it does not natively support chain or qubit loss events. We circumvent this limitation by first simulating conditional logical error probabilities and then using the law of total probability to compute the overall logical error rate. Specifically, by fixing a loss event $\loss(i,j)$, denoting the loss of the $i$-th chain at the $j$-th time step, we can efficiently simulate the conditional probability $\mathbb{P}(\text{logical error}| \loss(i,j))$. Indeed, simulating $\mathbb{P}(\text{logical error}| \loss(i,j))$ only requires two modifications to the original syndrome extraction circuit (which does not include chain loss or beacon qubits): (1) Add idling noise to unmeasured qubits during the measurements of beacon qubits; (2) At time step $j$, apply an independent single-qubit Pauli error with probability $0.75$ to every qubit in the $i$-th chain~\footnote{A Pauli error with probability $0.75$ (i.e., applying X, Y, Z, or I with $1/4$ probability each) sends any qubit to the maximally-mixed state, regardless of its original state.}. Based on this modified circuit, Stim compiles a new detector error model, which the BP-OSD decoder uses as its Tanner graph. This approach generalizes, allowing us to simulate conditional error probabilities for two or more loss events, such as $\mathbb{P}(\text{logical error}| \loss(i_1,j_1) \text{~and~} \loss(i_2,j_2))$.

Our simulation runs 6 rounds of syndrome extraction for the $[[72,12,6]]$ BB code. In this circuit, there are $N=13,824$ possible loss events $\loss(i,j)$, where $i$ is the chain index and $j$ is the time step. The probability of exactly $k$ chain loss events occurring during the entire syndrome extraction circuit is given by $p_k = \binom{N}{k} \ploss^k (1-\ploss)^{N-k}$. Under a pessimistic assumption $\mathbb{P}(\text{logical error}| k\text{ chain losses})=1$ for all $k\geq 3$, the overall logical error rate can be approximated by the upper bound
\begin{equation}\label{eq: approximation}
\begin{aligned}
\mathbb{P}(\text{logical error})&\lesssim p_0\mathbb{P}(\text{logical error}| \text{0 chain losses}) \\
&+p_1\mathbb{P}(\text{logical error}| 1\text{ chain loss}) \\
&+p_2\mathbb{P}(\text{logical error}| 2\text{ chain losses}) \\
&+ (1-p_0-p_1-p_2) .
\end{aligned}
\end{equation}
The $\mathbb{P}(\text{logical error}| \text{0 chain losses})$ term is obtained by simulating the original syndrome extraction circuit without chain loss or beacon qubits. Simulating the 1-loss term, $p_1\mathbb{P}(\text{logical error}| 1\text{ chain loss})=\ploss(1-\ploss)^{N-1}\sum_{(i,j)}\mathbb{P}(\text{logical error}| \loss(i,j))$, would require generating a separate circuit for all $N=13,824$ possible loss events $\loss(i,j)$. Due to resource constraints, we approximate this by randomly sampling 600 pairs of $(i,j)$ and simulating their respective conditional probabilities, $\mathbb{P}(\text{logical error}| \loss(i,j))$. We then estimate the full sum $\sum_{(i,j)}\mathbb{P}(\text{logical error}| \loss(i,j))$ as $N$ times the average conditional error probability of these 600 samples. A similar sampling approach is used for the third term, which is estimated from the average of 600 random double-loss samples of the form $\mathbb{P}(\text{logical error}| \loss(i_1,j_1) \text{~and~} \loss(i_2,j_2))$.

\section{Simulation results under a pessimistic beacon qubit measurement model}
\label{sect:pessimistic_beacon}

\begin{figure}
\includegraphics[width=0.85\linewidth]{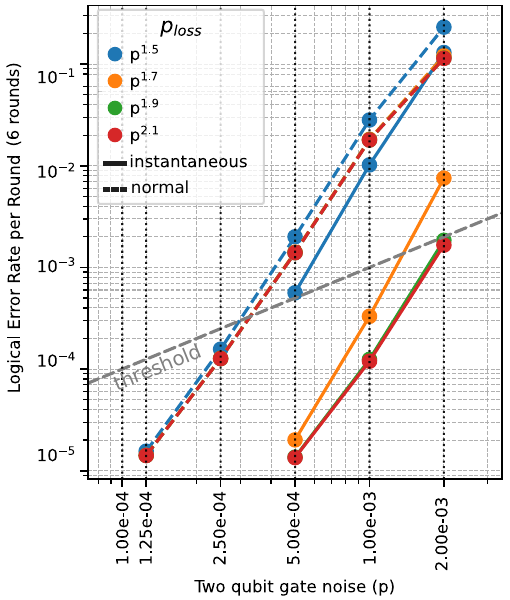}
\caption{Logical error rate as a function of $p$, with $\ploss = p^\alpha$ for $\alpha \in \{1.5, 1.7, 1.9, 2.1\}$. We plot dashed curves only for $\alpha \in \{1.5, 2.1\}$, as the 4 dashed curves are closely clustered.}
\label{fig: logical error rate physical error}
\end{figure}


In \cref{fig: logical error rate physical error}, we plot the logical error rate as a function of the 2-qubit gate error rate $p$ with $\ploss=p^{\alpha}$ for fixed $\alpha$.
We simulate two settings: (1) The solid curve (``instantaneous") represents beacon qubit measurements that take 0 time. This models an ideal scenario of free and perfect loss detection, while other measurements still take time according to the noise model. (2) The dashed curve (``normal") represents standard beacon qubit measurements, which are assumed to take the same amount of time as other measurements. In both settings, all beacon qubits are measured after every time step. This allows for the immediate detection of a chain loss, which is then transformed into an erasure by replacing the lost qubits with fresh ones in the maximally-mixed state.

In \cref{fig: logical error rate physical error}, the solid curves for $\alpha=1.9$ and $2.1$ are nearly identical, indicating that for $\alpha \ge 1.9$, chain loss events constitute a negligible part of the overall logical error rate, as further decreasing the chain loss probability yields very little improvement. In contrast, decreasing $\alpha$ from $1.9$ to $1.5$ causes a degradation of nearly two orders of magnitude, showing that chain loss becomes the dominant source of logical errors in this regime. This is also reflected in the threshold (the break-even point where the logical error rate per round equals $p$): for $\alpha \ge 1.9$, the threshold is roughly $2\cdot 10^{-3}$, while for $\alpha=1.5$, it degrades to $5\cdot 10^{-4}$. The gap between the dashed and solid curves highlights the benefit of fast beacon qubit measurements. For $\ploss=p^{1.5}$, the two curves differ only by a factor of about 3, indicating that chain loss detection based on beacon qubits is nearly as effective as free and perfect loss detection for large $\ploss$. However, for $\ploss=p^{2.1}$, the curves differ by a factor of more than 100. This large gap shows that fast beacon qubit measurements become significantly more valuable when $\ploss$ is small.


\section{Effect of single chain loss}
\label{sect:single chain loss}

\begin{figure*}
    \centering
    \includegraphics[width=0.8\textwidth]{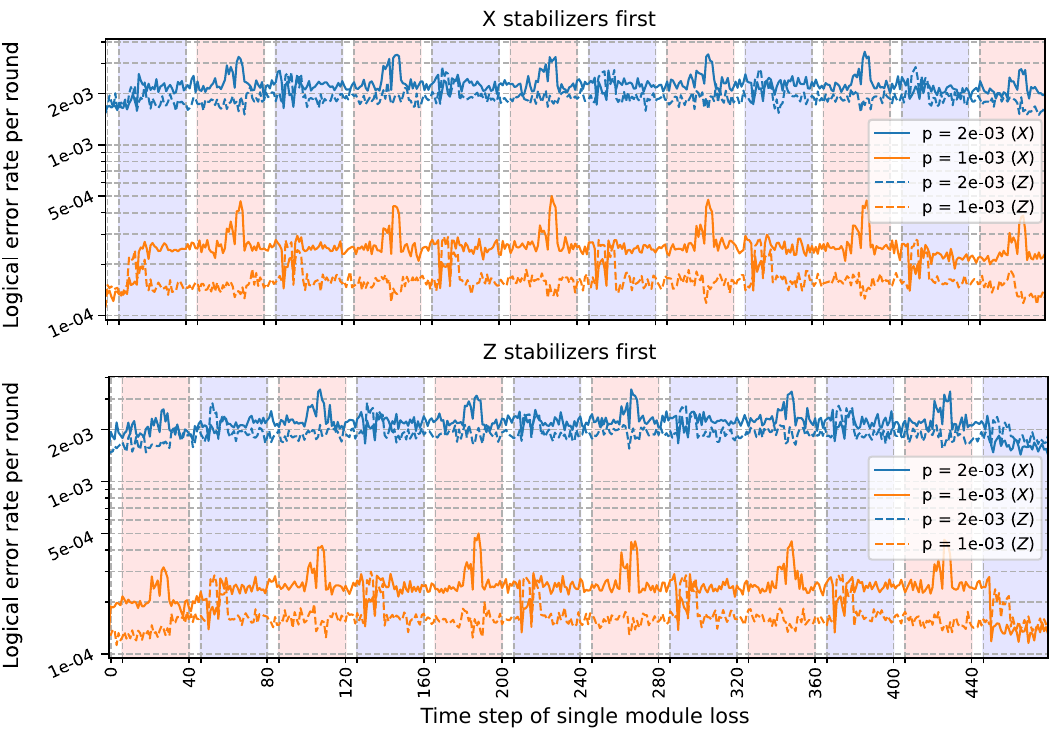}
    \caption{Effect of losing data module 0 at different time steps. The top plot shows results for the X-then-Z stabilizer measurement order, while the bottom plot uses the reverse Z-then-X order. Shaded areas (blue for X, red for Z) indicate stabilizer measurement steps, during which some modules are merged to execute 2-qubit gates. White areas represent ancilla reset/measurement steps, during which all modules are separate. Solid curves represent the $X$ logical error rate, and dashed curves represent the $Z$ logical error rate. $p$ in the legend denotes the physical 2-qubit gate error rate.}
    \label{fig: first chain lost}
\end{figure*}

\begin{figure*}
    \centering
    \includegraphics[width=0.8\textwidth]{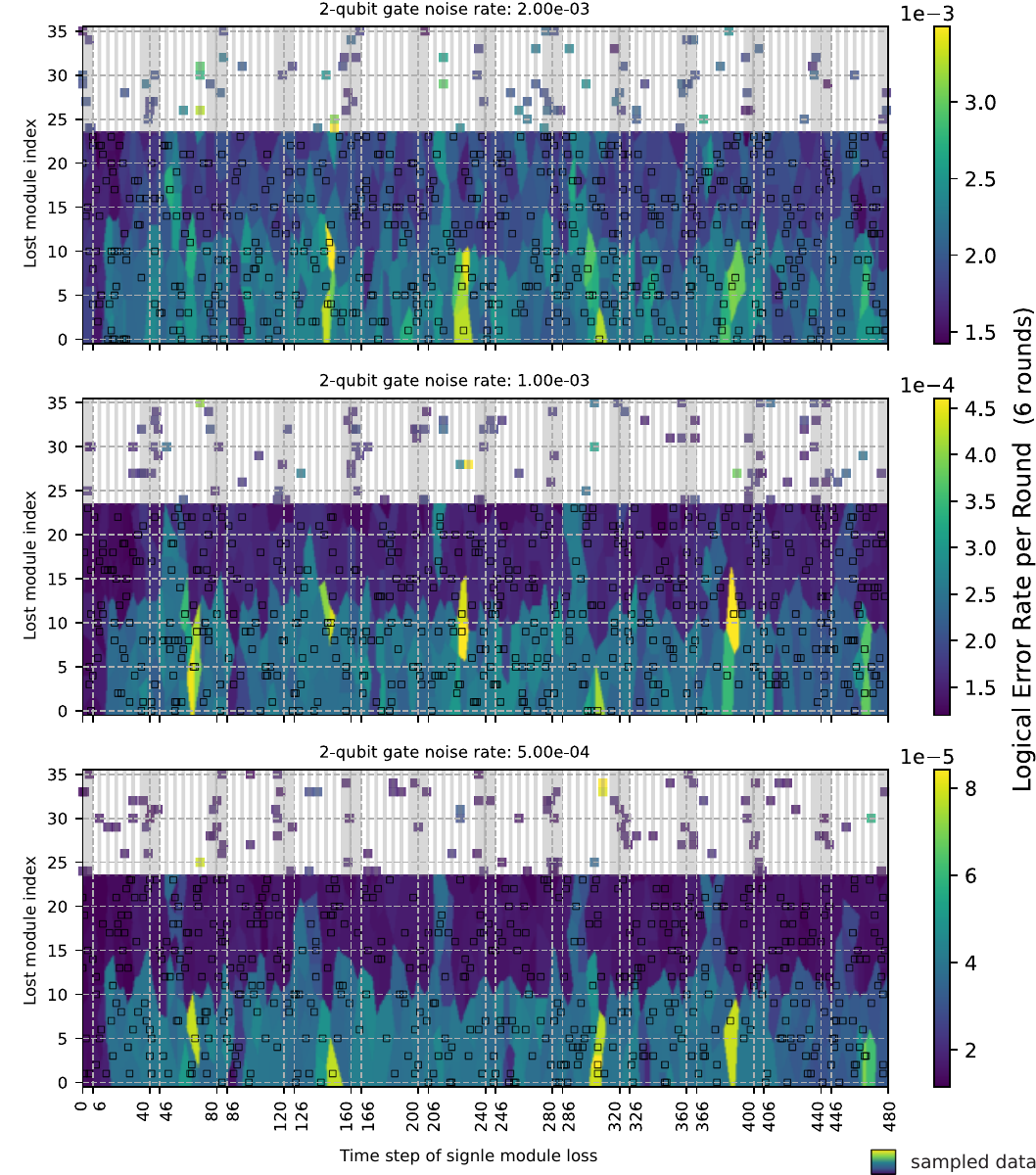}
    \caption{Effect of a single random chain loss occurring at a random time step. Each of the three plots corresponds to a different 2-qubit gate error rate. The colored squares show the logical error rate for $600$ randomly sampled chain loss events, defined by the time step ($x$-axis) and the lost module index ($y$-axis). During merged steps (unshaded regions), an ancilla module cannot be lost independently; it is lost jointly with the data module it is merged with. For this reason, no independent loss events are sampled for ancilla modules (indices 24-35) in the unshaded regions. To aid interpretation, the plot uses nearest-neighbor interpolation for unsampled data module (indices 0-23) loss events. Therefore, for the data modules, the color map represents true simulated data only at the locations of the colored squares.
    }
    \label{fig: one loss}
\end{figure*}

We investigate the effect of single chain loss within the 6-round, 480-timestep syndrome extraction circuit for the $[[72,12,6]]$ BB code. The system comprises 36 active modules: 24 data modules (indexed 0-23) and 12 ancilla modules (indexed 24-35). Note that while 24 ancilla modules exist in total (12 for X and 12 for Z), only 12 are active at any given time, as X and Z stabilizer measurements never overlap in the circuit.
The number of independent chain modules (potential loss targets) varies throughout the circuit. At time steps for operations like ancilla preparation or measurement, all 36 modules are separate. During two-qubit gate operations, the 12 ancilla modules merge with 12 data modules, resulting in 12 merged modules and 12 un-merged data modules. Summing the number of available modules at each of the 480 time steps, we find the total number of possible single chain loss events $\loss(i,j)$ is $N=13,824$.
All plots in this appendix (\cref{fig: first chain lost} and \cref{fig: one loss}) assume ``instantaneous" beacon qubit measurements (i.e., they take 0 time).

We first examine the effect of losing only data module 0 at different time steps in the circuit, with the results shown in \cref{fig: first chain lost}. As previously mentioned, this data module alternates between being merged with an ancilla module and standing alone. Our simulation assumes that if the loss occurs while the module is merged, the entire merged module (all 3 data and 3 ancilla qubits) is erased to the maximally-mixed state. Conversely, if the loss occurs during an unmerged step, only the 3 data qubits are erased.

In \cref{fig: first chain lost}, the $X$ logical error rate spikes if the data module is lost during $Z$ stabilizer measurements, while the $Z$ logical error rate spikes if the loss occurs during $X$ stabilizer measurements. This phenomenon is independent of the measurement order. As \cref{fig: first chain lost} illustrates, the top plot ($X$-then-$Z$ order) and the bottom plot ($Z$-then-$X$ order) exhibit the same spiking behavior in their respective logical error rates. We also note that the $X$ logical error rate is consistently slightly higher than the $Z$ logical error rate across all four simulated circuits (2 values of $p$ and both $X/Z$ measurement orders). 

In \cref{fig: one loss}, we simulate the effect of a single, random chain module loss occurring at a random time step. We assume the loss probability is independent of module size, meaning a merged data/ancilla module is just as likely to be lost as an unmerged one. We observe that the logical error rate decreases exponentially as the 2-qubit gate error rate decreases.

\end{document}